**Title:** Multiscale Modeling Primer: Focus on Chromatin and Epigenetics
Running title: multiscale chromatin modeling

**Authors:** Achal Mahajan[†1], Erik J. Navarro[†1], William Poole[†1], Carlos F Lopez[*2]

[1]Institute of Computation, Altos Labs Inc.
[2] Institute of Technology, Altos Labs Inc.

[†]These authors contributed equally to the manuscript
*Corresponding author - clopez@altoslabs.com

# Abstract

Essential life processes take place across multiple space and time scales in living organisms but understanding their mechanistic interactions remains an ongoing challenge. Advanced multiscale modeling techniques are providing new opportunities and insights into these complex processes. In cells, meters of chromatin are folded into a nucleus with a diameter on the order of microns. The three-dimensional chromatin structure coupled with biochemical processes that turn genes on or off, specify a given cell type through a complicated set of interactions collectively referred to as "epigenetics". Important epigenetic processes include the differential accessibility of genomic loci to transcription factors and chemical modifications to DNA and DNA-binding molecules such as histones. The dynamics of these epigenetic processes span timescales from milliseconds to years. How do chemical modifications consisting of a handful of atoms cooperate to modulate genome folding at the scale of the nucleus and impact organism outcomes? In this review, we highlight the inherently multiscale nature of chromatin organization, with a focus on computational modeling to bridge the gaps in our understanding of biochemical processes across scales. We review relevant chromatin biology, including major types of epigenetic modifications as well as the higher order chromatin structures to present a multiscale view of chromatin. We also review relevant computational methods to simulate chromatin structure, function, and dynamics, as well as experimental techniques that inform and validate said models. Finally, we argue that multiscale modeling provides a path forward towards understanding emergent behavior in this inherently multiscale system.

# Section I: Introduction

Fundamental processes for maintaining homeostasis and function in living organisms take place across multiple scales, ranging from atomic and molecular interactions to macroscale actions such as running or flying (1). The actions of an organism originate from atomic and molecular interactions, while dysfunctions, such as disease or aging, can often be traced back to failures at these same levels (2). To put these scales in context, the time scale gap between a

nanosecond – the time scale of bond shifts, and a second – the time scale of snapping fingers, is comparable to that between a second and a century. Therefore a central question in biology is how molecular interactions drive macroscale changes and, conversely, how macroscale stressors affect molecular interactions in higher organisms.

Our theoretical understanding of biological processes is typically limited by analytical, theoretical, and modeling tools, which describe spatiotemporal relationships within specific scales of space and time. For example, at the nanometer and nanosecond scales, bond formation and molecular dynamics play a fundamental role in DNA structure (Fig 1A, bottom left) (3). It is well established that even a single nucleotide mismatch can cause mutations that propagate in a deleterious manner throughout the whole organism (4). At the micrometer and microsecond resolution, DNA is organized into histones (Fig 1A, lower left), protein and DNA assemblies that play a central role in gene accessibility and DNA storage (5). At the micrometer to millimeter and microsecond to millisecond scales (Fig1A, mid upper right) we find the rich and complex mosaic of intracellular processes, with the added complexity that more measurement breadth comes at a cost of knowledge depth and vice versa. Beyond the millisecond and millimeter scales we now have features such as tissue organization, which gives way to whole organisms (Fig 1A, top right) (6). Some salient features are apparent from this description of organisms across scales. At atomic and molecular scales we have detailed knowledge about the components, but lack a comprehensive understanding of how they work together, and as we increase the scale metric, we can have a better understanding of the whole, but at the expense of losing information about individual components.

An interesting challenge emerges when we go off the space-time diagonal. In the large spatial but low temporal scales, organs such as human brains exhibit high complexity not only due to their cellular components, but also due to the fact that fundamental processes central to all aspects of the organism take place across time scales ranging from microseconds to the organism lifetime (7). On the small space and high time scales, we have complex systems such as chromatin, with highly relevant atomic interactions which can affect cells and the whole organism for time scales from hours to years (8).

In this review, we want to introduce basic concepts of modeling across multiple scales to biologists, while introducing basic biology concepts to quantitative scientists, with the goal of seeding connections between quantitative and biological researchers. We present a knowledge survey of the relevant spatiotemporal scales for living systems, with a specific focus on chromatin as a central component of all living organisms. For example, chromatin structures have been linked to cellular states, and epigenetic changes in chromatin have been linked to cellular health, disease, and aging (8). Chromatin poses an immense challenge from a scientific perspective due to the fact that structures and dynamics across multiple scales are relevant for its function and outcomes. In what follows we begin by introducing readers to basic concepts about chromatin biochemistry and some key features of chromatin that enable its function. This is followed by a section that introduces readers to some of the most important experimental and modeling tools used to study chromatin at different space and time scales, with the pedagogical goal to enable readers to associate a system of interest with appropriate experimental or modeling techniques.

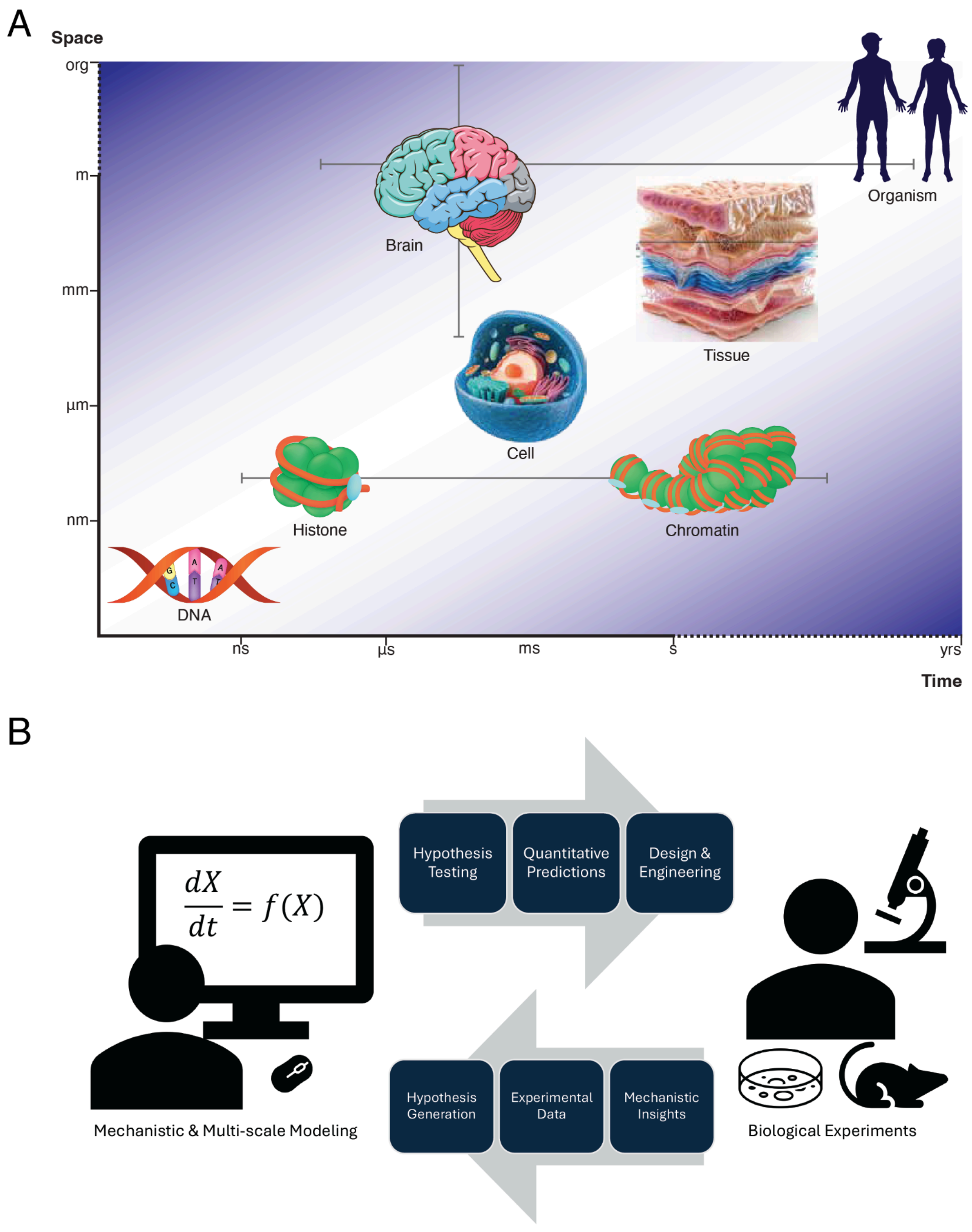

Figure 1. (A) A schematic representation of space and time scales relevant to chromatin structure and function, from atomistic space/time scales (bottom left) to whole organism (top right). (B) The bidirectional interaction between modelers (left) and experimentalists (right) to explain, predict, and validate observations.

This is followed by a section that introduces readers to some of the most important experimental and modeling tools used to study chromatin at different space and time scales, with the pedagogical goal to enable readers to associate a system of interest with appropriate experimental or modeling techniques. The following section introduces readers from both quantitative and biological sciences to modeling methodologies that use various forms of experimental data to explain and predict chromatin structure, function, and dynamics. Finally, we present some challenges and emerging solutions to bridge spatiotemporal gaps in the study of chromatin, which we also believe are generalizable to other biological processes. We note that although we exemplify the use of modeling techniques motivated by chromatin, the techniques are equally applicable to other biological complex systems.

| **Glossary** | |
|---|---|
| **Term** | **Description** |
| DNA | Single stranded DNA is composed of a linear polymeric chain of nucleic acid bases which bind together via base-pairing interactions into a double-helix consisting of double stranded DNA. |
| Histones | Proteins which bind together into an octamer protein complex around which DNA is wrapped. This histone octamer is called a nucleosome. |
| Nucleosome | Multimeric protein complex made of DNA wrapped around a histone octamer. |
| Chromatin | An organisms' DNA and the associated molecules bound to it which dictate its 3D and chemical properties, such as histones. Chromatin is frequently classified as euchromatin (accessible) and heterochromatin (less accessible). |
| Transcription Factors | A DNA-binding protein that regulates the transcription of one or more genes. |
| Enhancers | A non-coding region of DNA which serves as a regulatory element by recruiting transcription factors and other molecules. |
| Epigenetic Chemical Modifications | Chemical modifications of DNA and the nucleosomes bound to it including DNA methylation, histone acetylation, and histone methylation. |
| Epigenetic Readers and Writers | Proteins that bind to DNA or histones and chemically modify their substrate. Common modifications include methylation, acetylation, and phosphorylation. |
| Higher Order Chromatin Structures | These structures refer to the spatial and functional organization of the chromatin. Loops form when DNA is bent allowing distant locations along the sequence to colocalize. Topologically Associated Domains (TADs) and chromosomal territories are larger regions of densely interacting DNA. |
| Sequencing Based Measurements | These techniques use biochemical techniques to measure various aspects of chromatin structure by sequencing DNA. ATACseq measures the accessibility of chromatin to particular enzymes. Hi-C identifies sequences along the genome that are colocalized. CHIPseq uses antibodies to detect where DNA is interacting with the protein of interest. |
| Microscopy Based Measurements | Techniques which use microscopes in order to directly image the chromatin. Atomic Force Microscopy (AFM) uses a cantilever coupled to a laser to scan |
| | surfaces with high accuracy. Fluorescence Resonance Energy Transfer (FRET) measures the energy transfer between fluorophores to produce high spatial resolution measurements of spatially localized molecules. Cryo-EM is an electron microscopy technique where samples are flash frozen to preserve their native state. |
| Physics Based Models | Models that simulate the laws of physics appropriate in order to make quantitative predictions about how that system will behave. |
| Rule Based Models | Models that use heuristic rules to represent the behaviors of a system. These models are well suited for qualitatively understanding how lower scale behaviors give rise to higher scale phenomena. |

## Section II: Chromatin at multiple space and time scales

### DNA:

Deoxyribonucleic acid (DNA) is the most important information bearing molecule in cellular life. DNA typically takes the form of a double stranded helix consisting of four nucleic acids (known as base pairs), adenine (A), guanine (G), thymine (T), and cytosine (C), which are ordered in a sequence by a phosphate-sugar backbone. Each strand of DNA in the helix is a complement of the other with A binding to T and G binding to C. The genetic code is closely related to the central dogma of molecular biology where sequences of nucleotides are transcribed into ribonucleic acids (RNAs) which in turn are translated into proteins. Specifically, sets of three nucleic acids, called codons, are translated into one of about 20 amino acids that are the building blocks of proteins. The scale of DNA is immense; humans have ~3 billion base pairs encoding ~20,000 proteins and their vast amounts of regulatory information.

### Histones:

Histones are an example of a non-specific DNA binding protein, meaning that they can bind to almost any DNA sequence, though chemical modifications to DNA can affect histone binding (9,10). At the atomic and molecular level, DNA typically wraps around histone octamers, namely H2A, H2B, H3, and H4 proteins, forming nucleosomes. Nucleosomes are disc-shaped regions of tightly packed DNA of 1-4 nm in height and 13-22 nm in diameter. Additionally, H1 linker histones mediate the connections between nucleosomes. A strong association between DNA and histone proteins is correlated with decreased gene accessibility and gene expression (10–12).

The core proteins in a histone can also be replaced by so-called histone variants. These proteins exhibit modifications relative to the canonical histone proteins (H2A, H2B, H3, H4) and consequently can also modulate the DNA and histone affinity, the ability for nucleosomes to aggregate into multi-nucleosome complexes (9,10), and the recruitment of proteins that further modify the chromatin (9,10). Recent work has implicated that certain histone variants function as short-term epigenetic memory, temporarily maintaining chromatin state while the epigenetic landscape is reestablished after being diluted during cell division (13).

### Nucleosomes:

Nucleosomes consist of DNA wrapped around histone octamers (5). X-ray crystallography has provided an atomistic level view of the nucleosomes in which a stable complex is formed between an octamer of histones wrapped 1.7 times by a 147 - bp, left-handed DNA superhelix at a resolution of 1-2 Angstroms (14). These structures are stabilized by interactions between the histone core, the unstructured amino acid histone tails, and coiled DNA (14). The interactions are either electrostatic between the positively charged amino acid and negatively charged DNA backbone or between protein residues and DNA bases (15). Most of the bottom-up atomistic methods have used insight from the structural biology experiments to simulate the dynamics of a single nucleosome (~500,000 atoms) or at most 2 nucleosomes (~ 700,000 atoms) (16) with a simulation run time of up to 15 ms (16). However, current experimental techniques cannot achieve the atomistic resolution required to fully resolve the spatio-temporal dynamics of nucleosomes as at these timescales.

### DNA binding proteins:

Numerous proteins and protein complexes bind to DNA in order to modulate gene expression, chromatin structure, and cellular identity. Broadly, these can be classified into specific and non-specific DNA binding proteins. Non-specific binders may bind to any region of DNA regardless of its sequence. For example, High Mobility Group (HMG) proteins play important roles in chromatin organization (17). In contrast, specific DNA-binding proteins recognize and preferentially bind to specific motifs (sets of sequences) (18). Many of these proteins contain specific DNA binding domains such as zinc finger domains (19), helix-turn-helix domains (20), and leucine zipper domains (21). Some common examples of sequence specific DNA binding proteins include RNA polymerases,which transcribe DNA into RNA, and transcription factors, which activate and inhibit transcription by RNA polymerases. In the context of regulating gene expression, the specific DNA sequences which proteins bind to are known as cis-regulatory elements and include both promoters where RNA polymerase binds and to more distant regions, called enhancers and silencers, which activate and repress gene expression, respectively (22). Indeed, the ability of distant regions to modulate expression of genes located thousands or tens of thousands of base pairs away is evidence of complex structural organization of the chromatin. This includes the formation of loops that allow these regions and their associated binding proteins to interact with promoters (23). Indeed, it is likely that many sequence specific DNA-binding proteins function by manipulating the 3D organization of chromatin in order to modulate gene expression, and vice versa (24).

## Major types of epigenetic chemical modifications

Unlike the structural modifications mentioned above which are governed by the binding of various proteins and RNA to DNA in order to form complex molecular structures, epigenetic chemical modifications take the form of the addition or removal of various chemical groups to DNA or chromatin binding molecules, particularly histones (25).These modifications can be thought of as an extension of the genetic code with a huge number of possible combinations of marks potentially possible. We highlight some of the most important modifications below.

### DNA Methylation:

The most commonly studied chromatin epigenetic modification is DNA methylation, characterized by addition of a methyl group to cytosine bases in the DNA sequence form 5-methylcytosine (26). This modification is primarily found in somatic cells and is predominantly associated with symmetrical methylation at CpG sites, which contain one or more repeating pairs of cytosine adjacent to a guanine in the 5’ to 3’ direction (27). DNA methylation typically results in the silencing of genes in this region of the DNA, with significant implications for cancer and aging (28).

A number of enzymes called DNA methyltransferases are able to add and remove these modifications in a dynamic manner (29). Importantly, different cell types, such as those found in distinct tissues, exhibit different methylation patterns. Mechanistically, DNA methylation has been observed to regulate gene expression and cell differentiation (30). Embryonic stem cells (ES) exhibit a unique pattern, with a significant portion of 5-methylcytosine occurring in non-CpG contexts. The majority of the CpG sites undergo significant methylation throughout the genomic. One important exception are CpG islands, regions with densely clustered CpG sites that are situated in germ-line tissues and close to promoters of normal somatic cells, which often remain unmethylated (27). This unmethylated state facilitates gene expression (27). Conversely, methylation of a CpG island within a gene's promoter region leads to gene repression by impeding the transcriptional machinery from accessing DNA and initiating transcription (31).

### Histone Modifications:

Nucleosomes play an important role in epigenetic regulation because associated histone proteins can be chemically modified by enzymes. These chemical modifications on the histone proteins have physical consequences: they modulate the ability of nucleosomes to aggregate into multi-nucleosome complexes (32), and recruit proteins to chromatin (33). Some examples of well studied histone modifications include: histone methylation, where a methyl group is added primarily at the arginine and lysine residues (9,10); Histone acetylation, where an acetyl group is added primarily at lysine residues; and histone phosphorylation where a phosphoryl group is added at serine, threonine, or tyrosine residues (9). These modifications are not unique in the sense that multiple residues in a histone may be modified. Similarly, these modifications are not exclusive, enabling a combinatorially complex number of possible histone states. As of 2010, there were 24 sites on the canonical histones which were known to be methylated (34). Each site can be un-, mono-, di-, or tri-methylated. Assuming modifications are independent of each other, this gives an upper bound of $\sim 4^{24}$ on possible combinations of histone methylation patterns on a single nucleosome. Additionally, there is evidence that histone modifications interact with DNA methylation (35). For example, nucleosomes within CpG islands often contain histones with modifications involved in regulating gene expression (35,36).

## Higher Order Chromatin Structures as a Consequence of Epigenetics

### Loops:

As you zoom out from its molecular structure, chromatin is further organized into higher order structures called loops (37). These are formed when two non adjacent sections of chromatin are held together in close proximity. Loops can be formed by chromatin-chromatin binding interactions, such as those imposed by epigenetic modifications, by non-chromatin binding proteins like HP1 which can bridge two non-adjacent chromatin segments, and even through active loop extrusion by motor proteins like condensins and cohesins (38). Importantly, loop formation can have regulatory effects on genes, such as distant enhancers recruiting transcription factors to promoters via loop formation (24).

### Topologically Associated Domains (TADS):

As you zoom out further, structures consisting of sets of interacting loops come together to form Topologically Associated Domains (TADs) (39). These domains represent large regions of closely interacting sequences, genes, and other chromosomal features (39). Physical proximity facilitates biochemical reactions by localizing molecules to specific regions of the chromatin. TADs are typically defined based upon chromatin interactions measured via Hi-C and related experimental modalities. TADs are formed during development and are cell type specific (39). Disruption of TADs are known to be associated with many diseases such as cancer, brain disorders, etc (40).

### Heterochromatin and Euchromatin:

Chromatin is often categorized based upon the density of various epigenetic marks (41). These densities are related to the accessibility and compaction of the chromatin. Regions with large amounts of H3K9me3 marks are called *constitutive heterochromatin* and genes in these regions are typically repressed (42). The H3K27me3 modification marks *facultative heterochromatin* consisting of repressed genes that may be turned on in certain cellular processes (43). *Euchromatin* is more accessible and contains many highly transcribed genes (44). We comment that euchromatin and heterochromatin are sometimes called “A” and “B” based upon the interaction frequency in Hi-C maps (41).

### Chromosome Territories:

At the scale of the nucleus, chromosomes are organized into territories whereby each chromosome occupies a distinct subcompartment within the nucleus (45). It is believed that chromosome territories exist to prevent topological problems related to polymer mixing, such as knotting and interlocking, which are particularly important in the context of cell division where the chromosomes have to be carefully segregated into the daughter cells (46–48). Epigenetics marks, particularly H3k9me2 and H3k9me3, associated with nuclear envelope interactions have been shown to be important in maintaining chromosome territories, highlighting our point that modifications consisting of tens to hundreds of atoms have consequences at the scale of the nucleus (49,50).

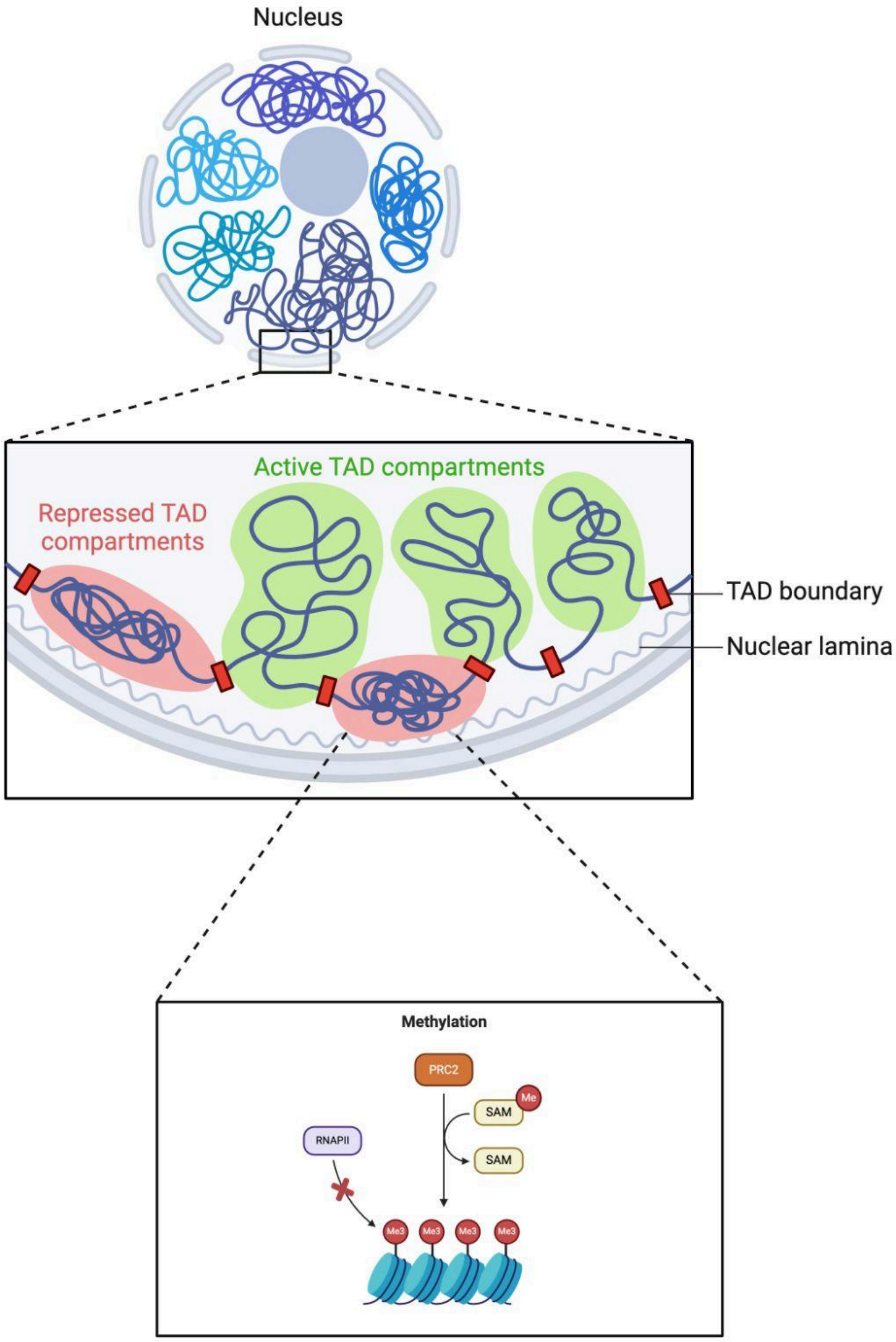

Figure 2. Chemical modifications consisting of a handful of atoms modulate chromatin organization across scales. Figure showing how chemical tags at the scale of DNA/nucleosomes have consequences at all multiple scales of chromatin organization: from topologically associated domains at the kilobase scale, to chromosome territories at whole genome scales. Created in BioRender (149).

## Chromatin defines cell state and identity

All cells in the human body share the same genetic material, yet human cells differentiate into at least 200 different cell types to perform specific functions (51). How a cell type specific transcriptional program is achieved has been the focus of intense research for the past several decades. Our current understanding is that a cell type specific patterning of epigenetic marks appears during cellular differentiation (52,53). This cell type specific epigenetic patterning causes a cell type specific transcriptional program through the ability of epigenetic marks to influence things like chromatin structure and dynamics, transcription factor recruitment, nuclear positioning, etc (33). Perturbing cellular state through techniques such as cellular reprogramming is accompanied by dramatic changes in the epigenome, a result which is consistent with the role of epigenetics in cell state specific transcriptional orchestration (54–56).

# Section III: Tools/methods used at different scales

In this section, we describe several commonly used methods for simulating chromatin at various scales.

| Model | Illustration | Description |
|---|---|---|
| Agent Based<br>Cellular Scale | 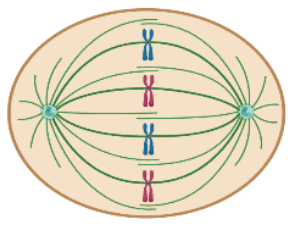 | Molecular state *s(t)* simulated as an update rule that can involve multiple numerical methodologies.<br>$s_{t+1} = F(s_t, t)$ |
| Network<br>Pathway Scale | 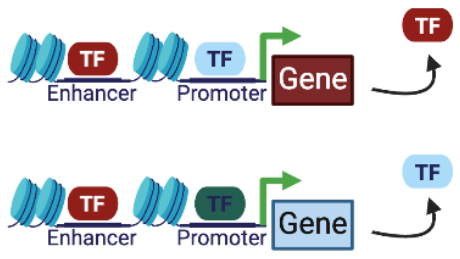 | Molecular state *s(t)* simulated as an ordinary differential equation or as a stochastic process (shown below).<br>$\frac{d\mathbb{P}(s)}{dt} = \sum_{s'} K(s' \to s)\mathbb{P}(s') - K(s \to s')\mathbb{P}(s)$ |
| Coarse-Grained Molecular<br>Multiple Nucleosome Scale | 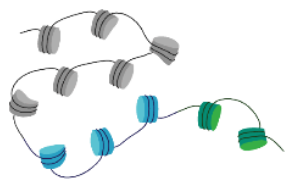 | 3D conformation *x(t)* simulated with Langevin dynamics which models the solvent implicitly.<br>$m\frac{d^2x}{dt^2} = -\nabla U(x) - \xi\frac{dx}{dt} + \sqrt{2m\xi k_B T}\,\mathcal{N}(t)$ |
| Atomistic Molecular<br>Single Nucleosome Scale | 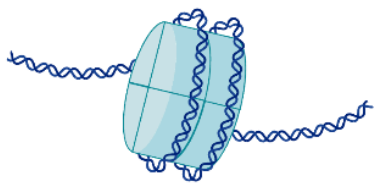 | 3D conformation *x(t)* simulated with Newtonian dynamics which models the solvent explicitly.<br>$m\frac{d^2x}{dt^2} = -\nabla U(x)$ |

Figure 3. A table of modeling modalities (left), illustration of the biological scale (center), and a description of the underlying mathematics (right).

### Atomistic Molecular Dynamics Simulations:

Atomistic molecular dynamics (MD) simulations directly model the interactions between individual atoms. The scale of these simulations ranges from thousands to millions of atoms and the simulated time scale spans from nanoseconds to microseconds. Importantly, atomistic MD simulations explicitly include the solvent molecules, such as water and ions, which typically make up the majority of the molecules simulated and use a bulk of the computational resources (57–59) . From a modeling perspective, atomistic MD involves solving Newton's equations of motion as shown in the lowest row of Figure 2. Here *m* is a vector of particle masses, *x* is a vector of particle positions, and $\nabla$ *U(x)* is the gradient of the system's energy also known as the force (60,61). In short, this equation is a sophisticated way of writing force equals mass times acceleration. More specifically, in most atomistic MD models, the energy function includes a Lennard Jones potential modeling excluded volume interactions (i.e. the fact that the physical location of two atoms cannot overlap), Van Der Waals forces, and an electrostatics term (60,61). Although conceptually simple, atomistic MD simulations are numerically challenging due to the huge numbers of atoms being simulated. State of the art computational methods require the use of highly optimized numerical methods, GPUs and ML/AI extensions to enable large MD simulations (62–65).

In the context of modeling chromatin, atomistic MD has provided significant insights into the structure of the nucleosome and specific effects of histone tails, DNA variations, and solvent/ion effects (66–68). However, performing multi-nucleosome simulations accounting for every permutation of histone modifications, histone variants, and nucleosome-binding proteins using atomistic level methods becomes challenging due to the extensive computational requirements of large MD simulations (66,69).

### Coarse Grained Molecular Dynamics Simulations:

Coarse grained molecular dynamics may be used in order to simulate increasingly large and complex systems. Coarse graining refers to grouping multiple atoms together in order to lower the total number of individual particles in a given molecular system so that larger systems become computationally tractable (70). One of the most common coarse graining techniques is to replace explicit solvent molecules with implicit solvent molecules modeled as viscous drag coupled to a random force which approximates the effect of solvent-solute interactions resulting in so-called Langevin Dynamics (71). Additionally, multiple atoms may be grouped together into coarse grained particles (70). Mathematically, coarse grained simulations still take the form of a stochastic differential equation such as the one shown in the second-to-bottom row of Figure 2. In this equation, $m$ is a vector of particle masses, $x$ is a vector of particle positions, $\nabla U(x)$ is the gradient of the system's energy, $\xi$ is the viscous drag coefficient, $k_B$ is Boltzman's constant, $T$ is temperature, and $N(t)$ are i.i.d. samples from a normal distribution with mean 0 and variance $m/(2\,\xi\,k_B T\,\Delta t)$ where $\Delta t$ is the time step used for numerical integration (72,73). By reducing the number of particles in the simulation, coarse graining has the potential to dramatically speed up simulations . However, the energy function $U(x)$ is often only heuristically defined based upon prior knowledge of interactions in the system instead of directly modeling physical forces.

In the context of chromatin modeling, coarse grained molecular dynamics simulations have been used on systems ranging from several nucleosomes to entire chromosomes (74). Tools such as MARTINI group several atoms together in order to simulate larger numbers of nucleosomes (75). Coarse graining has been systematically increased to the point where entire nucleosomes or even multiple nucleosomes may be represented by single particles (76). A popular class of models, known as bead spring polymer models, recreate this structure by treating nucleosomes (or even groups of nucleosomes) as spherical beads connected by springs.

Coarse grained models require additional parameters in their energy functions to model complex features such as internucleosomal interactions and the length of linker DNA, which must be measured or calculated accurately for these simulations to be realistic (77). Popular models (58) treat 10-50kb of chromatin as a single bead, and are able to successfully recapitulate known aspects of chromatin structure and dynamics including: diffusion, phase separation, epigenetic spreading, and the formation of higher order structures including loops, topologically associated domains and A/B compartments (41).

### Network Models:

The preceding models all describe chromatin as a physical set of polymers located in three dimensional space with various degrees of spatial resolution. However, simplified state based models can be more computationally efficient when the full 3D chromatin structure is not the primary object of investigation. For example, there is considerable evidence that chromatin structure can affect gene expression programs (78,79). Modeling such effects at a systems level requires simultaneously modeling the chromatin, transcription, translation, and feedback whereby genetic programs affect the chromatin structure (80,81). One approach is to coarse grain an entire 3D conformation of the chromatin into a single entity described as an abstract chemical species. Changes in the chromatin structure are represented by transitions between these abstract species. For example, $C_{Open} \rightarrow C_{Loop}$ might represent a transition where a segment of the chromatin forms an internal loop. In turn, $C_{open}$ may allow for a higher transcription of some gene X than $C_{Loop}$. Similarly, epigenetic modifications can be represented as discrete changes in the state of abstract chemical species e.g. $N \rightarrow N_M$ may represent the methylation of a nucleosome. These systems level models are commonly represented as chemical reaction networks (CRNs) which can be simulated stochastically as continuous time Markov chains using the Chemical Master Equation shown in the second row of Figure 2 (82). Here, *s* is an abstract state of the system, *P(s)* is the probability of a state *s* and *K(s → s')* is the transition rate from state *s* to *s'*. Note that this kind of system can also be simulated deterministically as ordinary differential equations which is appropriate for understanding the net dynamics of many cells (83). However, for understanding chromatin dynamics at the single cell level, stochastic simulations are likely to be more appropriate due to the intrinsic noise at low molecular abundances (83). Several frameworks have been developed for modeling epigenetics coupled to gene expression including the effect of histone modifications on reprogramming and methylation on cell fate decisions (84,85). Such models provide detailed mechanistic views of complex dynamical processes that cannot be easily assessed experimentally.

### Agent Based Models:

Agent based models take an increasingly abstract approach and allow the state, *s,* of a system to be updated based upon any function *F*. Unlike the preceding models, agent based models are not necessarily physics-based because the function *F* allows for numerical updates, rules-based updates, or virtually anything that can be codified as executable computer code (86–88). This provides modelers greater flexibility in formulating diverse models quickly and can in principle allow larger systems to be simulated. On the flip side, it is frequently harder to describe the details and underlying assumptions of agent based models due to their explicit flexibility (89,90). Therefore the validity of agent based models can be more difficult to gauge. In the context of chromatin modeling, agent based models have been used to model DNA replication in E. Coli (91) and have allowed for highly simplified cell-state based models at multicellular scales (88).

### Statistical Models:

Although this review is predominantly focused on mechanistic models of chromatin, it is still important to briefly touch on statistical and machine learning (ML) driven models. The success of ML to solve protein structures with algorithms such as AlphaFold (92) suggests that these kinds of techniques may play a role in understanding chromatin structure and dynamics as well. Indeed, AlphaFold also demonstrates how using an ML architecture that mirrors the mechanistic process of co-transcriptional folding can lead to better predictions (92). However, any statistical / AI driven approach is limited by the data available to train the model. At this time, there is very limited single-cell chromatin data and the data that does exist tends to be sparse and noisy and therefore cannot capture the entire 3D structure of the chromatin at the single molecule level (93). However, using techniques such as ATACseq and CHIPseq, researchers have been able to capture information on the locations and modifications of individual histones and transcription factors (94,95). At larger scales, Hi-C has allowed researchers to identify topologically associated domains (TADs) representing likely contact points between distant chromatin sequences (96). These data types have also been used as predictors of cell state, cell type, and transcriptomics profiles (97). Indeed, the ability of chromatin based measurements to statistically predict other molecular, cellular, and phenotypic information is a clear indication of the importance of understanding the mechanistic underpinnings of these processes.

In summary, MD based mechanistic modeling approaches describe the chromosome at diverse scales ranging from single atoms to strings of interacting nucleosomes but are challenging to scale. Network models focus on coupling specific conformational aspects of chromatin structure to other biochemical processes such as gene regulatory networks but rarely take a holistic view of chromatin structure. Agent based models have allowed abstract cellular states to be scaled to increasingly large systems, but similarly to network based models require massive simplifications. Finally, statistical models can identify structural features from data but are often difficult to convert to mechanistic and dynamical understanding. Indeed one focus of this review is to argue that coupling different modeling scales is necessary to answer questions related to how chromatin structure and dynamics affect cells, tissues, and ultimately organisms.

## Section IV: Data modalities specific to physical models

Many existing multiscale models use polymer and statistical physics to provide insight into the dynamics of chromatin organization across various scales. However, these methods often face difficulties in accurately reconstructing dynamics due to the challenge of identifying the correct set of parameters (98). To overcome this, combining physics-based and data-driven approaches is crucial for determining the necessary parameters of the dynamical model. In this section, we highlight the diverse range of data modalities that can be used to infer the parameters of the model.

### Atomic Force Microscopy (AFM):

Atomic force microscopy (AFM, also called scanning force microscopy) has become a widely used tool for both imaging and force measurement, particularly for studying nucleosomal arrays of varying lengths in isolated chromatin and DNA (99). AFM data can be supplemented in the dynamical models to understand the structural dynamics, interactions, and mechanical properties of chromatin.

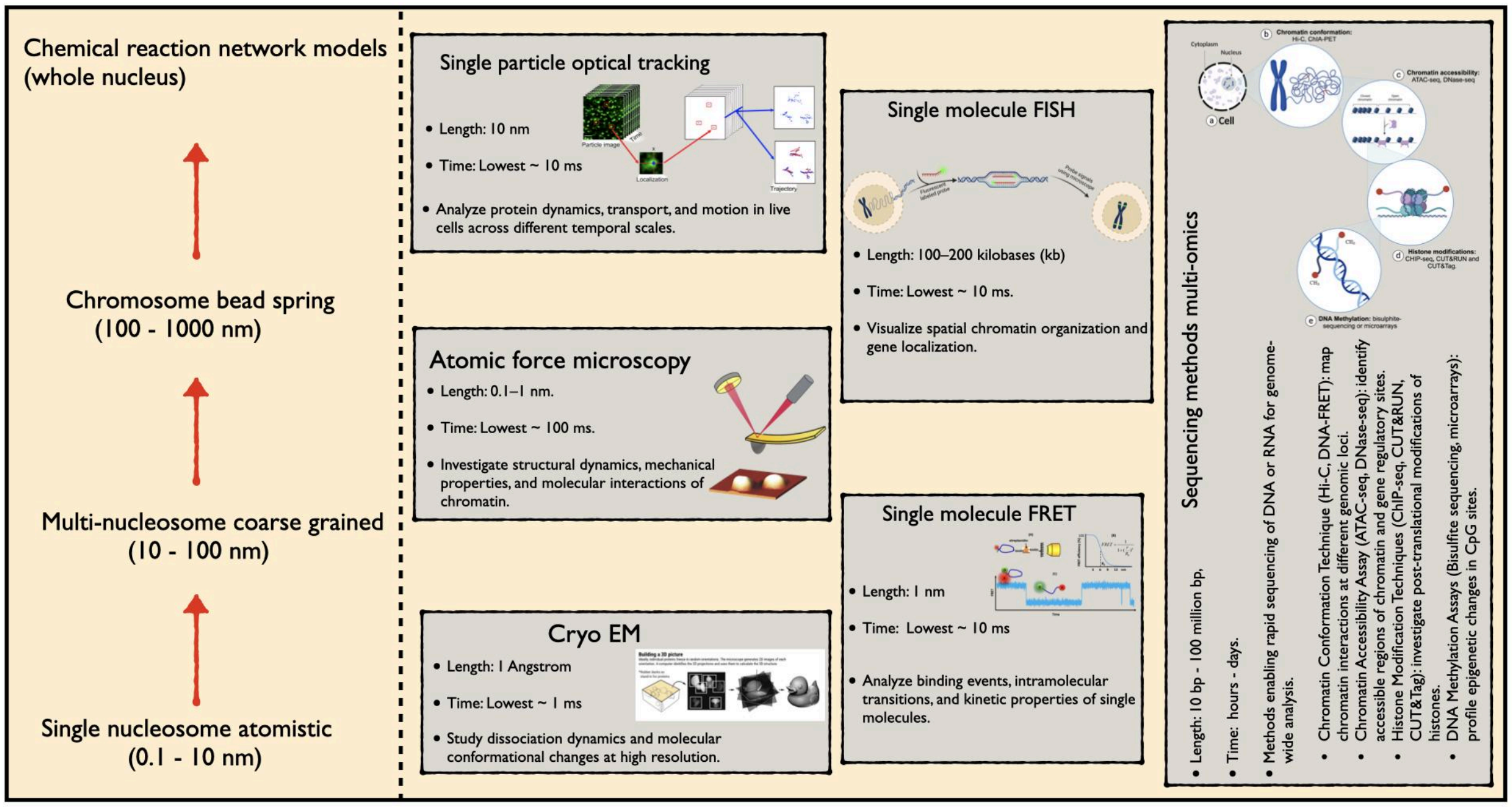

Figure 4. A multi-scale framework for measuring chromatin dynamics and genome organization, showcasing experimental measurements that can be integrated with modeling strategies to infer or refine model parameters. The left column represents modeling scales, ranging from atomistic single nucleosome models to whole-nucleus reaction networks. The right column highlights experimental methodologies across different scales  used to study chromatin dynamics, accessibility, and epigenetic modifications, providing data to inform and enhance these models.

With its high resolution imaging capabilities, AFM allows for the measurement of physical properties of chromatin such as diameter, and flexibility (99,100), which are essential for predicting how changes in these properties might influence accessibility of DNA to transcription factors and other regulatory proteins. Epigenetic modifications often involve the binding of proteins to DNA. AFM can be used to study these interactions, providing data that can be used to model how these proteins affect DNA structure and function. Additionally, AFM can be used to measure DNA-histone binding strength (101), nucleosomal positioning (102) and the actions of complexes that modify chromatin structure, such as ATP-dependent nucleosome remodeling complexes (103). As such, AFM serves as a powerful tool for the development of physics-based models of chromatin and epigenetic modifications.

### Cryogenic Electron Microscopy (cryoEM):

Another useful technique to obtain the structure of molecules at single angstrom resolution is Cryogenic electron microscopy (cryoEM) (104). Compared to traditional electron microscopy, cryoEM involves flash-freezing samples in liquid nitrogen before imaging, which preserves their native hydrated state (104). At single angstrom resolution, cryoEM is capable of resolving amino acid side-chains, providing a complete structural snapshot of proteins. Additionally, because cryoEM samples contain proteins in their native hydrated state, they often exhibit multiple conformations of the target proteins. This feature enables the inference about structural changes related to protein function (105). In the context of chromatin, cryoEM has been used to capture the structure of multi-nucleosome complexes. Notably, several distinct structures have been identified, including a flat two-start helix (106) and a stacked structure of telomeric nucleosomes (107). One can postulate that these different structures result from differences in epigenetic composition, but proving so will require paired data where the epigenetic composition of the target nucleosomes is known. Most studies involving cryoEM in epigenetics have focused on resolving the structure of epigenetic readers and writers bound to chromatin-nucleosome complexes (108), (109).

### Fluorescence Resonance Energy Transfer (FRET):

Fluorescence resonance energy transfer, or FRET, is a technique by which energy is transferred between fluorophores in a manner that is spatially dependent (110). This makes FRET extremely sensitive and useful for biology, as it is capable of accurately probing distances to single nanometer resolution: the resolution of protein-protein interactions. FRET has been used to measure nucleosome stability (111), distances between histones (112), and nucleosome-nucleic acid complexes (113). In the context of epigenetics, groups have used FRET to measure chromatin packing densities resulting from epigenetic modifications (114). Such measurements could be used to validate or parameterize epigenetic-based chromatin models that capture distances at the single to ten nanometer length scales.

### Fluorescence In Situ Hybridization (FISH):

Fluorescence in situ hybridization (FISH) is a technique that uses fluorescent probes to bind to specific DNA sequences in the genomes (115). It is used for visualizing and spatial detection of genetic material including specific genes (115). One way to incorporate FISH data into a model is to use the spatial distribution of various genomic elements as initial conditions. Computational models have also been used to validate observations made using FISH, such as those related to radial positioning of chromosomes in the nuclear volume of human and primate lymphocytes (45). A relationship between the gene density of a chromosome territory (CT) and its distance to the nuclear center was observed as a result of probabilistic global gene positioning code depending on CT sequence length and gene density (116).

FISH data can also be used to validate model predictions associated with epigenetic changes. For example, if a model predicts that a specific epigenetic modification will alter the spatial organization of a gene, this can be tested by performing FISH on cells with and without the modification and comparing the results. Alternatively, FISH can be used in conjunction with Hi-C to validate the spatial proximity of specific genomic regions that are expected to be in physical proximity based on Hi-C data. Most importantly, these measurements have begun to reveal how the 3D spatial organization of chromatin influences gene expression and its function (117).

## OMICS (Sequencing Based Methods):

The recent advancements in sequencing technologies have led to an explosion of methods for inferring information about chromatin structure. These techniques all make use of combining high throughput DNA sequencing with a variety of enzymatic modifications that allow specific features of the chromatin to be selected (118).

- ATAC-Seq (Assay for Transposase-Accessible Chromatin using sequencing) measures DNA accessibility using transposon insertions that primarily occur in open stretches of DNA without histones or other binding proteins to qualitatively measure DNA accessibility by computing the relative frequency of transposon insertions (119).
- Chip-Seq (chromatin immunoprecipitation sequencing) uses antibodies to pull down specific binding proteins (such as methylated histones or transcription factors) which have been cross-linked to the chromatin (95). Cut & Run (120) as well as Cut & Tag (121) are optimized versions of Chip-Seq commonly used in epigenetic research.
- Bisulfite sequencing chemically targets methylated nucleotides in order to measure sequence-specific DNA methylation (122).
- Hi-C and related techniques cross-link chromatin to itself and then sequence the cross-linked fragments in order to identify pairwise contacts in the chromatin structure. This results in large chromatin probability pairwise contact maps (123).

Jointly, these methods allow for a wealth of multiscale chromatin structural data. Most of these methods have been used predominantly on bulk samples resulting in population level chromatin information as opposed to single molecular information (124). However, recent developments such as scATACseq protocols are beginning to produce single-molecule information although noise and sparsity are still major issues when using single cell data (93).

Beyond measuring chromatin directly, the chromatin state can be partially inferred from transcriptomics measurements (125). In particular, single cell RNAseq is an increasingly common technique for capturing the transcriptomic profiles of single cells (126). Additionally, some techniques allow scRNAseq to be paired with single cell ATACseq measurements (127). This data is just beginning to be used to understand chromatin structure dictates gene expression profiles at a mechanistic level (128,129). Importantly, these techniques lend themselves well to time course experiments which offer the possibility of beginning to decipher the interplay between chromatin dynamics and gene regulation (10,22,81).

## Motivation for the use of physical models

Chromatin state and its activity are highly regulated across different scales. Physical structures have been identified in the chromatin spanning from associations between several nucleosomes to correlated chemical modifications of hundreds of kilobases, such as those observed in heterochromatin. Important structural patterns include large nuclear sub-compartments, typically around 1-10 megabases to more compact topologically associating domains (TADs), which are several hundreds of kilobases in size, and even smaller contact domains and loops. Additionally, these spatial features may change over diverse temporal scales ranging from hours to weeks resulting in highly multiscale dynamics.

Understanding the coupling between these diverse spatial and temporal scales requires state-of-the-art multiscale theoretical and computational approaches. The utility of multiscale chromatin modeling is amplified because of the inherent trade-offs in temporal and spatial resolution in most experimental techniques. On one end, high spatial resolution techniques like Cryo-EM and crystallography produce snapshots with no temporal resolution. On the other end, microscopy based techniques may have high temporal resolution but are limited in spatial resolution by diffraction (130) and the number of available fluorescent probes (131). 'Omics methods such as ATAC-seq and Hi-C offer a middle ground with the possibility of capturing multiple dynamical snapshots at an intermediate resolution. Multiscale modeling offers the possibility of bridging the gaps and unifying observations between these different experimental data modalities. In the following section, we will look at some of these and also lay out strategies to multi-scale modeling strategies specifically in context to diseases and how we can use them as a tool to predict changes in macroscopic cellular state due to disease and an intervention.

## Section V: Challenges and future perspectives

An understanding of chromatin structure and dynamics and its impact in living systems represents a frontier in understanding the fundamental processes governing cellular function, differentiation, and disease. Chromatin operates across an exceptional range of spatiotemporal scales, from atomic-level interactions to cellular level dynamic responses to organism-level organization across cell types. The goal of creating complex multiscale models of chromatin is a large undertaking that will require collaboration across the scientific community. We firmly believe that this undertaking will pay off in dividends by illuminating the mechanisms by which chromatin structure and function guides various biological processes. Chromatin organization has been associated with many diseases with associations bridging many scales (132–134). Additionally, there are many pharmaceutical compounds known to modulate chromatin which may have clinical relevance, such as histone deacetyl transferase inhibitors, histone acetyl transferases (135). However, it is rarely clear how empirical observations about chromatin structure, epigenetic marks, and DNA sequences drive disease in a mechanistic fashion. Multiscale models offer the potential to bridge this understanding by showing how a perturbation at one scale can cause a change in phenotype at a different scale (136). This understanding will pave the way for new therapeutics that target chromatin organization (137).

At the atomistic level, molecular modeling methods have provided fundamental insights about DNA structure, chromatin packing, nucleosome behaviors, and the potential impact of epigenetic modifications. However, these simulations are limited by the size of the system, the level of resolution, and the temporal domains in which they operate (138,139). Although coarse graining methods have greatly enhanced system size and time-scale capabilities, they come at a cost of resolution.

The mesoscale poses a further challenge in understanding chromatin structure and dynamics. The organization of chromatin into loops, topologically associating domains (TADs), and compartments, highlights the importance of spatial and temporal context in gene regulation.

Experimental techniques such as Hi-C, cryo-EM, and super-resolution microscopy have illuminated these structures, yet they have inherent limitations to capture the dynamic nature of these processes (140). Integrating static datasets with time-resolved data, such as single-molecule imaging or live-cell microscopy, could provide critical insights into the mechanisms underlying chromatin folding and remodeling (141).

At the organism scale, the main challenge in understanding the impact of chromatin dynamics lies on the heterogeneity of the cellular milieu in a tissue and the rich and complex interactions among cell types (142,143). This problem is exacerbated by the accompanying changes with developmental stage and organism age (144,145). Single-cell techniques such as single-cell Hi-C and ATAC-seq, have begun to address this cellular variability but remain limited by resolution, throughput, and noise (146–148).

New modeling insights have recently been applied to biological systems that may allow for increasingly complex and predictive models of chromatin to bridge scales and be compared to multiple experimental modalities. Our preferred approach is compositional modeling, where different modeling frameworks can be seamlessly integrated together in order to create complex multiscale simulations from sub-components that can be developed independently. This approach contrasts with static modeling methods with fixed methodologies at each scale. The Vivarium software package is an example of this approach and has already been applied to whole cell simulation which includes a minimal model of bacterial DNA and its coupling to other processes (136). By using a modular framework to connect different numerical methodologies, the scientific method can be applied to empirically test different methods of coupling these simulations. Similarly, multiple data modalities can be seamlessly used to initialize or validate each module independently. This allows for iterative, scalable, and reusable modeling frameworks that can be systematically built, tested, and improved based upon their predictive power.

Regardless of the computational strategy used to build multiscale chromatin models, success in this process will necessarily require intentional interdisciplinary collaboration between modelers, biologists, and software engineers. Modeling without careful consideration of the fundamental biology and available data risks missing key aspects of the underlying system and producing models which cannot be validated. Similarly, modeling without careful consideration of the scalability of the computational methods risks producing models incapable of capturing the essential complexities of chromatin and its dynamics.